\documentclass[twocolumn,showpacs,superscriptaddress,amssymb,10pt,pra,floatfix,aps,longbibliography]{revtex4-1}
\pdfoutput=1
\usepackage{graphicx}
\usepackage{amsmath}
\usepackage{dcolumn}
\usepackage{bm}
\usepackage{epsfig}
\usepackage{color}
\usepackage{longtable}
\usepackage{leftidx,amsmath}
\usepackage{natbib} 
\usepackage[utf8]{inputenc}
\usepackage{braket}
\usepackage{tikz}
\usepackage{makecell}
\usepackage[flushleft]{threeparttable}
\usepackage[utf8]{inputenc}

\DeclareMathOperator*{\SumInt}{%
\mathchoice%
  {\ooalign{$\displaystyle\sum$\cr\hidewidth$\displaystyle\int$\hidewidth\cr}}
  {\ooalign{\raisebox{.14\height}{\scalebox{.7}{$\textstyle\sum$}}\cr\hidewidth$\textstyle\int$\hidewidth\cr}}
  {\ooalign{\raisebox{.2\height}{\scalebox{.6}{$\scriptstyle\sum$}}\cr$\scriptstyle\int$\cr}}
  {\ooalign{\raisebox{.2\height}{\scalebox{.6}{$\scriptstyle\sum$}}\cr$\scriptstyle\int$\cr}}
}

\def\redme#1#2#3{ $ \left\langle #1 \left\Vert
	#2 \right\Vert #3 \right\rangle $ }
\def\redmem#1#2#3{  \left\langle #1 \left\Vert
	#2 \right\Vert #3 \right\rangle   }
\def\emc2{\ensuremath{~m_\text{e}c^2}}

\begin{document}
	\preprint{}
	%
%
%
%
	\title{Vacuum polarization and finite nuclear size effects in the two-photon decay of hydrogen-like ions}
%
%
%
%

	\author{J.~Sommerfeldt}
	\affiliation{Physikalisch--Technische Bundesanstalt, D--38116 Braunschweig, Germany}
	\affiliation{Technische Universit\"at Braunschweig, D--38106 Braunschweig, Germany}
	
	\author{R.~A.~M{\"u}ller}
	\affiliation{Physikalisch--Technische Bundesanstalt, D--38116 Braunschweig, Germany}
	\affiliation{Technische Universit\"at Braunschweig, D--38106 Braunschweig, Germany}

	\author{A.~V.~Volotka}
	\affiliation{Helmholtz Institute Jena, D--07743 Jena, Germany}
	
	\author{S. Fritzsche}
	\affiliation{Helmholtz Institute Jena, D--07743 Jena, Germany}
	\affiliation{Theoretisch-Physikalisches Institut, Friedrich-Schiller-Universität Jena, D-07743 Jena, Germany}
	
	\author{A.~Surzhykov}
	\affiliation{Physikalisch--Technische Bundesanstalt, D--38116 Braunschweig, Germany}
	\affiliation{Technische Universit\"at Braunschweig, D--38106 Braunschweig, Germany}	

	\date{\today \\[0.3cm]}

%
%
\begin{abstract}
The total two-photon decay rate of hydrogen-like ions is studied using relativistic quantum electrodynamics. In particular, we analyse how finite nuclear size and QED vacuum polarization corrections affect the decay rate. To calculate these corrections, a finite basis set method based on $B$-splines is used for the generation of quasi-complete atomic spectra and, hence, of the relativistic Green's function. By making use of this $B$-spline approach, high precision calculations have been performed for the $2s_{1/2} \to 1s_{1/2} + 2\gamma$ and $2p_{1/2} \to 1s_{1/2} + 2\gamma$ decay of hydrogen-like ions along the entire isoelectronic sequence. The results of these calculations show that both, QED and finite nuclear size effects, are comparatively weak for the $2s_{1/2} \to 1s_{1/2} + 2\gamma$ transition. In contrast, they are much more pronounced for the $2p_{1/2}\to 1s_{1/2} + 2\gamma$ decay, where, for hydrogen-like Uranium, the decay rate is reduced by 0.484\% due to the finite nuclear size and enhanced by 0.239\% if the vacuum polarization is taken into account.
\end{abstract}
\maketitle

\section{Introduction}
Theoretical investigations of two-photon transitions in hydrogen-like ions have a long history going back to the seminal work by Göppert-Mayer \cite{doi:10.1002/andp.19314010303}. In that work, solutions of the non-relativistic Schrödinger equation have been applied to calculate the rate of the $2s \to 1s$ two-photon transition in neutral Hydrogen. Four decades later, the first fully relativistic calculations were done to describe the two-photon decay of the meta-stable $2s_{1/2}$ state~\cite{PhysRevLett.29.1123, PhysRevA.24.183}. Being performed within the framework of second-order perturbation theory for the electron-photon coupling, these calculations are rather demanding. They require a representation of the entire atomic spectrum including the positive and negative continuum. Theoretical investigations of two-photon transitions in hydrogen-like ions, therefore, have become a testbed for the development of second-order computational approaches. Several of such methods have been developed during the last decades leading to more precise predictions of the decay rates~\cite{PhysRevA.40.1185, Amaro_2011, PhysRevA.93.012517}. The increased accuracy of the calculations has allowed to investigate how the total and differential rates are influenced by nuclear and even QED effects. In particular, finite nuclear size and mass corrections to the two-photon decay of $n=2$ hydrogenic states have been studied \cite{fried_center--mass_1963, PhysRevA.34.2871,PhysRevA.26.1142, Labzowsky_2005}. Furthermore, QED effects have been discussed to all orders in $\alpha Z$ for one-photon transitions \cite{PhysRevA.69.022113, PhysRevA.71.022503, volotka_radiative_2006} but only to the leading order for the two-photon decay of hydrogen-like ions~\cite{ KARSHENBOIM1997375, PhysRevA.69.052118}. 

Despite the recent interest in high-precision calculations of two-photon transitions in hydrogen-like ions, no systematic fully relativistic analysis of the finite nuclear size and QED corrections to the decay rates has been performed. To the best of our knowledge, the first steps towards this analysis were done by Parpia and Johnson~\cite{PhysRevA.26.1142} who have discussed the finite nuclear size corrections for the $2s_{1/2} \to 1s_{1/2} + 2\gamma$ transition in dipole approximation. Moreover, in the work by Labzowsky and co-authors~\cite{Labzowsky_2005}, calculations with a finite nucleus have been performed but with low relative precision. More accurate calculations of two-photon transitions in hydrogenic systems are required, however, to get a more complete picture of bound electronic states and as benchmark data for future second-order atomic calculations. 

In this contribution, therefore, we present a theoretical study of the finite nuclear size and QED vacuum polarization corrections to the two-photon decay rate of hydrogen-like ions. To analyse these effects, we employ the relativistic quantum electrodynamics approach whose basic equations are recalled in Sec.~\ref{QEDdescr}. In Sec.~\ref{finiteB}, we discuss the finite basis set method used the construct the second order transition matrix element. The implementation of this method requires knowledge about the electron-nucleus interaction potential. In Sec.~\ref{potentials}, we show how this potential can be modified from the pure Coulombic case to include finite nuclear size and vacuum polarization effects. Since we aim for a discussion of small effects, the accuracy of the two-photon calculations should be sufficiently high. In Sec.~\ref{CompDet}, we discuss how such accuracy can be achieved by using quadruple precision arithmetic. The results of our calculations are presented in Sec.~\ref{ResADis} for the $2s_{1/2} \to 1s_{1/2} + 2\gamma$ and $2p_{1/2} \to 1s_{1/2} + 2\gamma$ decay of hydrogen-like ions in the range from $Z=1$ to $Z=92$. The results of these calculations are in good agreement with previous predictions and indicate that the finite nuclear size and QED corrections to the decay rate are of opposite sign. We found that the finite nuclear size reduces the decay rates while the vacuum polarization enhances them. Both effects are more pronounced for the $2p_{1/2} \to 1s_{1/2} + 2\gamma$ transition compared to the $2s_{1/2} \to 1s_{1/2} + 2\gamma$ case. The summary of these results and outlook are finally given in Sec.~\ref{SumAOut}. Relativistic units $\hbar = m_e = c = 1$ are used throughout this paper if not stated otherwise. 

\section{Theory} \label{Theory}
\subsection{QED description of two-photon decay} \label{QEDdescr}

\begin{figure} [hb]
\begin{tikzpicture}
\node (s1) at (-0.025,  -0.3) {}; 
\node (s2) at (0.025,  -0.3) {}; 
\node (x1) at (0.0, -1.3) {}; 
\node (x2) at ( 0.0, -2.6) {}; 
\node (e1) at (-0.025, -3.6) {}; 
\node (e2) at (0.025, -3.6) {}; 

\draw[thick] (s1) -- (e1);
\draw[thick] (s2) -- (e2);
\fill[fill=black] (0.0,-2.9)--(-0.1, -3.15)--(0.1, -3.15);
\fill[fill=black] (0.0,-1.8)--(-0.1, -2.05)--(0.1, -2.05);
\fill[fill=black] (0.0,-0.7)--(-0.1, -0.95)--(0.1, -0.95);

\draw[color=black, fill=black] (x1) circle (.07);
\draw[color=black, fill=black]  (x2) circle (.07);

\draw (x1) arc(160:20:0.2);
\draw (0.375,-1.3) arc(200:340:0.2);
\draw (0.75,-1.3) arc(160:20:0.2);
\draw (1.125,-1.3) arc(200:250:0.2);

\draw (x2) arc(160:20:0.2);
\draw (0.375,-2.6) arc(200:340:0.2);
\draw (0.75,-2.6) arc(160:20:0.2);
\draw (1.125,-2.6) arc(200:250:0.2);

\node at (-0.4,-1.3) {$x_1$};
\node at (-0.4,-2.6) {$x_2$};
\node at (1.8,-1.3) {$k_1, \epsilon_1$};
\node at (1.8,-2.6) {$k_2, \epsilon_2$};
\node at (0.3,-0.3) {$f$};
\node at (0.3,-3.5) {$i$};
\node at (0.0,-3.9) {$S_1$};

\node (s12) at (-0.025+3.3,  -0.3) {}; 
\node (s22) at (0.025+3.3,  -0.3) {}; 
\node (x12) at (0.0+3.3, -1.3) {}; 
\node (x22) at (0.0+3.3, -2.6) {}; 
\node (e12) at (-0.025+3.3, -3.6) {}; 
\node (e22) at (0.025+3.3, -3.6) {}; 

\draw[thick] (s12) -- (e12);
\draw[thick] (s22) -- (e22);
\fill[fill=black] (0.0+3.3,-2.9)--(-0.1+3.3, -3.15)--(0.1+3.3, -3.15);
\fill[fill=black] (0.0+3.3,-1.8)--(-0.1+3.3, -2.05)--(0.1+3.3, -2.05);
\fill[fill=black] (0.0+3.3,-0.7)--(-0.1+3.3, -0.95)--(0.1+3.3, -0.95);

\draw[color=black, fill=black] (x12) circle (.07);
\draw[color=black, fill=black]  (x22) circle (.07);

\draw (x12) arc(160:20:0.2);
\draw (0.375+3.3,-1.3) arc(200:340:0.2);
\draw (0.75+3.3,-1.3) arc(160:20:0.2);
\draw (1.125+3.3,-1.3) arc(200:250:0.2);

\draw (x22) arc(160:20:0.2);
\draw (0.375+3.3,-2.6) arc(200:340:0.2);
\draw (0.75+3.3,-2.6) arc(160:20:0.2);
\draw (1.125+3.3,-2.6) arc(200:250:0.2);

\node at (-0.4+3.3,-1.3) {$x_1$};
\node at (-0.4+3.3,-2.6) {$x_2$};
\node at (1.8+3.3,-1.3) {$k_2, \epsilon_2$};
\node at (1.8+3.3,-2.6) {$k_1, \epsilon_1$};
\node at (0.3+3.3,-0.3) {$f$};
\node at (0.3+3.3,-3.5) {$i$};
\node at (0.0+3.3,-3.9) {$S_2$};
\end{tikzpicture}
\caption{Leading order Feynman graphs corresponding to the two-photon transition $\ket{i}\to \ket{f} + \gamma(k_1, \epsilon_1) + \gamma (k_2, \epsilon_2)$.} \label{Feynman}
\end{figure}
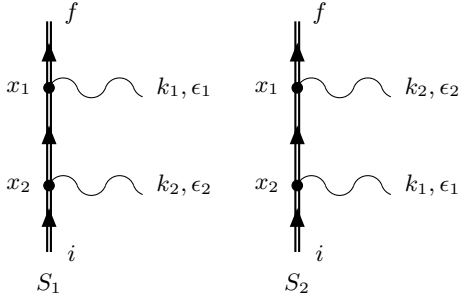

Within the framework of quantum electrodynamics, two-photon decay of hydrogenlike ions can be described, to the leading order, by the two Feynman diagrams presented in Fig. \ref{Feynman}. As usual, in these diagrams the wavy lines display photons emitted with wave vectors $k_1$, $k_2$ and polarization vectors $\epsilon_1$, $\epsilon_2$. Moreover, the double straight lines display the bound electron which proceeds from the initial $\ket{i} = \ket{n_i\kappa_i\mu_i}$ to the final $\ket{f} = \ket{n_f\kappa_f\mu_f}$ hydrogenic state. Here, $n$ is the principal quantum number, $\kappa$ is the Dirac quantum number and $\mu$ is the projection of the total angular momentum $j = \vert \kappa \vert - \frac{1}{2}$.

By using the Feynman correspondence rules, we can write the $S$ matrix element for each diagram from Fig. \ref{Feynman}. For example, the matrix element for the first diagram is given by

\begin{equation}
\begin{aligned}
S_1 = &(-ie)^2\int \text{d}^4x_1~ \int \text{d}^4x_2~ \bar{\phi}_f(x_1)\gamma^{\mu_1}A_{\mu_1}^* (x_1)\\
&\times S_F(x_1, x_2)\gamma^{\mu_2} \phi_i(x_2)A^*_{\mu_2}(x_2)
\end{aligned}
\end{equation}

\noindent where $S_F$ is the Feynman propagator

\begin{equation} \label{FProp}
S_F(x_1, x_2) = \frac{1}{2\pi i} \int_{-\infty}^\infty \text{d}w \SumInt_\nu \frac{\phi_\nu(\boldsymbol{x}_1) \bar{\phi_\nu}(\boldsymbol{x}_2)}{E_\nu + w(1+i\delta)} e^{iw(t_1-t_2)}~,
\end{equation}

\noindent and the $\phi_i$ and $\phi_f$ are solutions of the relativistic Dirac equation with corresponding eigenenergies $E_i$ and $E_f$~\cite{MOHR1998227}. Here, the summation over the intermediate states $\ket{\nu} = \ket{n_\nu\kappa_\nu\mu_\nu}$ is understood as a sum over all bound-states and an integral over the positive and negative continuum. 

By constructing the $S$ matrix element for the second Feynman diagram in a similar way, inserting the electron and photon wave functions and carrying out the integration over time and frequency explicitly we obtain

\begin{equation}
S_1 + S_2 = -\frac{4 \pi^2ie^2}{\sqrt{\omega_1\omega_2}} M_{fi}(\boldsymbol{k}_1,\boldsymbol{\epsilon}_1, \boldsymbol{k}_2,\boldsymbol{\epsilon}_2)
\end{equation}

\noindent where the matrix element $M_{fi}$ is given by

\begin{equation} \label{Matel1}
\begin{aligned}
M_{fi}(&\boldsymbol{k}_1,\boldsymbol{\epsilon}_1, \boldsymbol{k}_2,\boldsymbol{\epsilon}_2)=\\ &\sum_\nu \frac{\bra{f}\hat{R}^\dagger(\boldsymbol{k}_1,\boldsymbol{\epsilon}_1)\ket{\nu}\bra{\nu}\hat{R}^\dagger(\boldsymbol{k}_2,\boldsymbol{\epsilon}_2)\ket{i}}{E_\nu - E_i + \omega_2} \\
&+ \frac{\bra{f}\hat{R}^\dagger(\boldsymbol{k}_2,\boldsymbol{\epsilon}_2)\ket{\nu}\bra{\nu}\hat{R}^\dagger(\boldsymbol{k}_1,\boldsymbol{\epsilon}_1)\ket{i}}{E_\nu - E_i + \omega_1}~.
\end{aligned}
\end{equation}

\noindent Here, the wave and polarization vectors are now given as 3-vectors since we have integrated over time. This matrix element is the starting point from which we investigate all properties of two-photon decay. We can further evaluate it by writing the electron-photon interaction operator as

\begin{equation}
\hat{R}(\boldsymbol{k},\boldsymbol{\epsilon}) = \boldsymbol{\alpha}\cdot (\boldsymbol{\epsilon} + G \hat{\boldsymbol{k}})e^{i\boldsymbol{k}\boldsymbol{r}} - Ge^{i\boldsymbol{k}\boldsymbol{r}}
\end{equation}

\noindent where we introduced an arbitrary gauge parameter $G$. This parameter is later set to either $G = 0$ for the velocity gauge or $G = \sqrt{(L+1)/L}$ for the length gauge. It is convenient to decompose the interaction operator $\hat{R}$ into spherical tensors. For emission of a photon in the direction $\hat{\boldsymbol{k}} = (\theta, \phi)$, this expansion reads

\begin{equation}
\hat{R}(\boldsymbol{k},\boldsymbol{\epsilon}) = 4 \pi \sum_{pLM} i^{L - \vert p\vert} [\boldsymbol{\epsilon} \cdot \boldsymbol{Y}_{LM}^{(p)*}(\hat{\boldsymbol{k}})] a_{LM}^{(p)} ~.
\end{equation}

\noindent where $\boldsymbol{Y}_{LM}^{(p)*}$ are vector spherical harmonics and the index $p$ describes electric ($p = 1$), magnetic ($p = 0$) and longitudinal ($p = -1$) multipole fields  \cite{PhysRevA.24.183, Manakov_2001}. By inserting this decomposition into Eq.~(\ref{Matel1}) and using the Wigner-Eckart theorem we obtain

\begin{equation} \label{Matel2}
\begin{aligned}
&M_{fi}(\boldsymbol{k}_1,\boldsymbol{\epsilon}_1, \boldsymbol{k}_2,\boldsymbol{\epsilon}_2)= \sum_{p_1L_1M_1}\sum_{p_2L_2M_2}i^{\vert p_1\vert + \vert p_2 \vert - L_1 - L_2}\\
&\times[\boldsymbol{\epsilon_1}^* \cdot \boldsymbol{Y}_{L_1M_1}^{(p_1)}][\boldsymbol{\epsilon_2}^* \cdot \boldsymbol{Y}_{L_2M_2}^{(p_2)}] \widetilde{M}_{fi}(p_1L_1M_1, p_2L_2M_2)~,\\
\end{aligned}
\end{equation}

\noindent where the matrix element for a particular multipole transition $(p_1L_1,p_2L_2)$ is given by \nopagebreak
\begin{widetext}
\begin{equation} \label{Matel3}
\begin{aligned}
\widetilde{M}_{fi}(p_1L_1&M_1, p_2L_2M_2) = \sum_{n_\nu\kappa_\nu \mu_\nu}  \frac{(4\pi)^2}{\sqrt{(2j_i+1)(2j_\nu+1)}} \\
&\times \Bigg[\langle j_f \mu_f L_1 M_1 \vert j_\nu \mu_\nu \rangle \langle j_\nu \mu_\nu L_2 M_2 \vert j_i \mu_i \rangle \frac{\redmem{n_f \kappa_f}{a_{L_1}^{(p_1)\dagger}}{n_\nu\kappa_\nu}\redmem{n_\nu \kappa_\nu}{a_{L_2}^{(p_2)\dagger}}{n_i\kappa_i}}{E_\nu - E_i + \omega_2}\\
&+ \langle j_f \mu_f L_2 M_2 \vert j_\nu \mu_\nu \rangle \langle j_\nu \mu_\nu L_1 M_1 \vert j_i \mu_i \rangle\frac{\redmem{n_f \kappa_f}{a_{L_2}^{(p_2)\dagger}}{n_\nu\kappa_\nu}\redmem{n_\nu\kappa_\nu}{a_{L_1}^{(p_1)\dagger}}{n_i\kappa_i}}{E_\nu - E_i + \omega_1}\Bigg]\\
\end{aligned}
\end{equation}
\end{widetext}

\noindent and with \redme{n_f\kappa_f}{a_{L}^{(p)\dagger}}{n_i\kappa_i} being the reduced one-photon matrix element \cite{10.5555/1215645, Grant_1974}.

With the help of matrix element (\ref{Matel2})-(\ref{Matel3}), we can now calculate observable quantities. For example, the triple differential (in energy and angles) decay rate is given by

\begin{equation} \label{AngleDiffRate}
\begin{aligned}
\text{d}&W(\omega_1, \Omega_1, \Omega_2, \boldsymbol{\epsilon}_1, \boldsymbol{\epsilon}_2) = e^4 \frac{\omega_1 (E_i - E_f- \omega_1)}{(2\pi)^3} \frac{1}{2j_i+1}  \\
&\sum_{\mu_i \mu_f} \left\vert M_{fi}(\boldsymbol{k}_1,\boldsymbol{\epsilon}_1, \boldsymbol{k}_2,\boldsymbol{\epsilon}_2) \right\vert^2 \text{d}\Omega_1 \text{d}\Omega_2 \text{d}\omega_1~.
\end{aligned}
\end{equation}

\noindent Here, we assume that the magnetic sublevels of both, the initial and final state, remain unresolved and hence sum over $\mu_f$ and average over $\mu_i$. If, moreover, the direction and polarization of the emitted photons are not detected in a particular study, we can obtain the energy-differential rate:

\begin{equation} \label{eDiffRate}
\frac{\text{d}W}{\text{d}\omega_1} = \sum_{\boldsymbol{\epsilon}_1\boldsymbol{\epsilon}_2} \int \frac{\text{d}W}{\text{d}\Omega_1\text{d}\Omega_2\text{d}\omega_1} \text{d}\Omega_1\text{d}\Omega_2~.
\end{equation}

\noindent The integration over the photon angles and the summation over their polarization can be easily performed using Eq. (\ref{Matel2}). Namely, by inserting the matrix element $M_{fi}(\boldsymbol{k}_1,\boldsymbol{\epsilon}_1, \boldsymbol{k}_2,\boldsymbol{\epsilon}_2)$ into Eqs. (\ref{AngleDiffRate}) and (\ref{eDiffRate}) and using the orthonormality relations

\begin{equation}
\sum_\epsilon \int \text{d}\Omega [\boldsymbol{\epsilon}^* \cdot \boldsymbol{Y}_{LM}^{(p)}][\boldsymbol{\epsilon} \cdot \boldsymbol{Y}_{L'M'}^{(p')*}] = \delta_{pp'}\delta_{LL'}\delta_{MM'}~,
\end{equation}

\noindent we obtain

\begin{equation} \label{eDiff2}
\begin{aligned}
d&W(\omega_1) = \sum_{p_1L_1M_1}\sum_{p_2L_2M_2} \text{d}W_{p_1L_1p_2L_2}(\omega_1) ~.
\end{aligned}
\end{equation}

\noindent As seen from this expression, the single differential decay rate can be obtained as the sum of individual multipole contributions 

\begin{equation} \label{rateMult}
\begin{aligned}
\text{d}W_{p_1L_1p_2L_2}(\omega_1) =  e^4 \frac{\omega_1 (E_i - E_f- \omega_1)}{(2\pi)^3}\frac{1}{2j_i+1} \text{d}\omega_1\\
\times \sum_{\mu_i \mu_f} \left\vert \widetilde{M}_{fi}(p_1, L_1, M_1, p_2, L_2, M_2) \right\vert^2
\end{aligned}
\end{equation}

\noindent with no interference terms. Finally, by integrating Eq. (\ref{eDiff2}) over the frequency $\omega_1$, we obtain the total decay rate

\begin{equation} \label{totalRate}
W = \frac{1}{2} \int_0^{E_i-E_f} \frac{\text{d}W(\omega_1)}{\text{d}\omega_1} \text{d}\omega_1~,
\end{equation}

\noindent where the factor $1/2$ is introduced to avoid double photon counting, see \cite{Labzowsky_2005} for more details.

\subsection{Finite basis expansion} \label{finiteB}

As seen from Eq. (\ref{Matel3}), the evaluation of the two-photon decay rate requires to perform a summation over all intermediate states $\ket{n_\nu \kappa_\nu \mu_\nu}$ explicitly. Being infinite and running over both, bound and continuum states, this summation is not a simple task. In the past, a large number of methods has been proposed to perform this summation. For example the Coulomb Green's function approach \cite{Swainson_1991,MOHR1998227,KOVAL2003191} and various finite basis set methods \cite{Amaro_2011,PhysRevA.93.012517,PhysRevA.40.1185} were both successfully applied to perform second-order calculations. In this work, we will also use finite basis sets constructed from $B$-spline functions. Since $B$-spline sets in atomic physics have been discussed in the literature \cite{PhysRevLett.93.130405,Bachau_2001}, we restrict ourselves to some basic formulas.

We start our analysis from the usual Dirac equation for the electron in the field of the nucleus. Assuming the potential for the interaction is spherically symmetric, the radial part of this equation can be given by

\begin{equation}
H_{\kappa} \phi_{n\kappa}(r) = E_{n\kappa} \phi_{n\kappa}(r)~,
\end{equation}

\noindent where

\begin{equation} \label{DiracH}
H_\kappa = \left(\begin{array}{cc}
V(r) + 1 & -\frac{\text{d}}{\text{d}r} + \frac{\kappa}{r} \\
\frac{\text{d}}{\text{d}r} + \frac{\kappa}{r} & V(r) - 1
\end{array}\right)
\end{equation}

\noindent is the radial Dirac Hamiltonian and

\begin{equation}
\phi_{n\kappa}(r) = \frac{1}{r} \left(\begin{array}{c}
G_{n\kappa}(r)\\
F_{n\kappa}(r)
\end{array}\right)
\end{equation}

\noindent are the radial wave functions. To solve this equation, we approximate these wave functions by

\begin{equation} \label{waveFunctions}
\phi_{n\kappa}(r) = \sum_{i=1}^{2N} c^i_{n\kappa} u^i_{\kappa}(r)~,
\end{equation}

\noindent where the $u^i_{\kappa}(r)$ are square-integrable, linearly independent two-component functions that satisfy proper boundary conditions. In order to find the expansion coefficients $c^i_{n\kappa}$, one usually applies the principle of least action $\delta S = 0$ with

\begin{equation}
S = \bra{\phi_{n\kappa}} H_\kappa \ket{\phi_{n\kappa}} - E \langle \phi_{n\kappa} \vert \phi_{n\kappa} \rangle ~.
\end{equation}

\noindent Inserting in this expression the wave functions (\ref{waveFunctions}), the variational condition reduces to the generalized eigenvalue problem

\begin{equation} \label{EigenValue}
\begin{aligned}
K^{ik}_{\kappa}c^k_{n\kappa} = E_{n\kappa} B^{ik}_{\kappa}c^k_{n\kappa}~,\\
\end{aligned}
\end{equation}

\noindent where

\begin{subequations} \label{matrices}
\begin{eqnarray}
K^{ik}_{\kappa} &=& (\bra{u^i_{\kappa}}H_\kappa\ket{u^k_{\kappa}} + \bra{u^k_{\kappa}}H_\kappa\ket{u^i_{\kappa}})/2~,\\
B^{ik}_{\kappa} &=& \langle u^i_{\kappa} \vert u^k_{\kappa} \rangle~.
\end{eqnarray}
\end{subequations}

\noindent In order to calculate the $K^{ik}_{\kappa}$ and $B^{ik}_{\kappa}$ matrix elements, to solve the generalized eigenvalue problem (\ref{EigenValue}) and find the $c^i_{n\kappa}$ coefficients, we have to specify the functions $u^i_\kappa$. Following the method by Shabaev and co-workers \cite{PhysRevLett.93.130405}, we choose 

\begin{subequations} \label{baisFct}
\begin{eqnarray}
u^i_{\kappa}(r) &=& \left(\begin{array}{c}
\pi^i(r)\\
\frac{1}{2} (\frac{\text{d}}{\text{d}r} + \frac{\kappa}{r})\pi^i(r)
\end{array}\right), \hspace{1cm} i \leq N\\
u^i_{\kappa}(r) &=& \left(\begin{array}{c}
\frac{1}{2} (\frac{\text{d}}{\text{d}r} - \frac{\kappa}{r})\pi^{i-N}(r)\\
\pi^{i-N}(r)
\end{array}\right), \hspace{0.65cm} i > N
\end{eqnarray}
\end{subequations}

\noindent where $\{\pi^i(r)\}_{i=1}^N$ is the set of $B$-splines on the finite interval $(0,R_\text{cav})$. The radius $R_\text{cav}$ of the finite cavity, in which the ion is enclosed, is chosen large enough to ensure that the wave functions (\ref{waveFunctions}) are a good approximation for the real hydrogenic wave functions of the initial and final state. Here, the first and last $B$-splines on the interval are omitted to fulfil the proper boundary conditions~\cite{PhysRevLett.93.130405}: $F_{n\kappa}(0) = 0$ for $\kappa < 0$, $G_{n\kappa}(0) = 0$ for $\kappa > 0$ and $F_{n\kappa}(R_\text{cav}) = G_{n\kappa}(R_\text{cav}) = 0$. This particular choice of the functions $u^i_\kappa$, by construction (\ref{baisFct}), avoids the unphysical so-called spurious states which often show up in finite basis methods and slow down the convergence of any calculation.

\subsection{Electron-nucleus interaction potential} \label{potentials}

In the previous section, we have shown how the hydrogenic Dirac wave functions can be constructed in the framework of a finite basis method. As seen from Eqs. (\ref{DiracH}), (\ref{EigenValue}) and (\ref{matrices}), the practical implementation of this method requires knowledge about the electron nucleus interaction potential. Generally, this potential can be obtained by

\begin{equation}
V(r) = e\int \text{d}^3\boldsymbol{r}' \frac{\rho(r')}{\vert r - r' \vert}~,
\end{equation}

\noindent where $\rho(r')$ is the nuclear charge density. In the present study, we will consider a few models of $\rho(r')$ to better understand the nuclear effects on the two-photon decay rate. The most naive approach is the point-like nucleus represented by the charge distribution

\begin{equation}
\rho_\text{pnt}(\boldsymbol{r}) = \rho_0 \delta(\boldsymbol{r})~,
\end{equation}

\noindent which leads to the usual Coulomb potential $V = \alpha Z/r$. 

In order to account for the effects of the finite nuclear size, we will either model the nucleus as a homogeneously charged sphere

\begin{equation} \label{SphereModel}
\rho_\text{sph}(\boldsymbol{r}) = \rho_0 \Theta(R_\text{sph} - r)~,
\end{equation}

\noindent where $R_\text{sph}$ is related to the root-mean
square charge radius by $R_\text{sph} = \sqrt{5/3}R$, or as a Fermi distribution

\begin{equation} \label{FermiModel}
\rho_\text{Fermi}(\boldsymbol{r}) = \frac{\rho_0}{1 + \exp[(r - c)/a]}~.
\end{equation}

\noindent For the latter $a = 2.3/(4 \ln 3)$ fm and the parameter c is given by the approximate formula

\begin{equation}
c \approx \frac{5}{3} R^2 - \frac{7}{3}a^2\pi^2~,
\end{equation}

\noindent see \cite{johnson_atomic_2007,PhysRevA.83.012507}. For both nuclear models, (\ref{SphereModel}) and (\ref{FermiModel}), the nuclear radius is taken from reference~\cite{ANGELI201369}.

Apart from the finite nuclear size effects, other phenomena can also influence the two-photon decay rate of hydrogen-like ions. For example, the interaction of the electron with the quantum vacuum may affect the wavefunctions \cite{PhysRevA.92.042510, PhysRevA.101.032511}, the energy levels \cite{PhysRevA.83.012507}, and the transition operators \cite{PhysRevA.69.022113, PhysRevA.71.022503, volotka_radiative_2006} and, hence, the decay rate. In the one-loop approximation, two QED contributions are usually considered: vacuum polarization and self energy. While the complete treatment of these two corrections to the second-order amplitude is a very complicated task, the vacuum polarization to the leading order $\alpha (\alpha Z)^2$ can be described by an effective Uehling potential. The corresponding diagrams can be obtained by vacuum-polarization insertions in the initial and final wavefunctions and the propagator between the two photon emissions, cf. Fig. \ref{Feynman}. The diagrams, which can not be accounted for by an effective potential, such as photon-vacuum-polarization-loop or two-photon-vacuum-polarization-loop diagrams, vanish in the free-loop approximation, and, therefore, contribute beyond the leading order. Thus, the leading-order contributions of diagrams with the ordinary vacuum-polarization insertions can be treated by solving the Dirac equation with the Uehling potential

\nopagebreak
\begin{equation} \label{UehlPot}
\begin{aligned}
V_\text{Uehl}(r) &= -\alpha Z \frac{2\alpha}{3\pi} \int_0^\infty \text{d}r'~4\pi r' \rho(r')\\ 
&\times \int_1^\infty \text{d}t \left(1 + \frac{1}{2t^2}\right) \frac{\sqrt{t^2-1}}{t^2}\\
&\times \frac{\exp (-2\vert r-r'\vert t) - \exp (-2\vert r+r'\vert t)}{4rt}~,
\end{aligned}
\end{equation}

\noindent where the nuclear charge distribution $\rho(r')$ is assumed to be spherically symmetric. To account for both, the vacuum polarization and the finite nuclear size effects, we sum $V_\text{Uehl}$ and the potential resulting from Eq. (\ref{SphereModel}) or (\ref{FermiModel}), respectively. Technical details of these calculations will be discussed in the next section.

\section{Computational Details} \label{CompDet}
The present work aims to investigate the finite nuclear size and QED effects on the two-photon decay rates of hydrogen-like ions. These effects are rather small requiring a high accuracy of our calculations. Namely, their numerical uncertainty should be remarkably lower compared to the size of the discussed effects. As usual in atomic structure calculations, we verify the accuracy of the computations by testing the gauge invariance of the decay rates. Following our discussion above, calculations within the so-called velocity and length gauge are compared. For all results presented in the next chapter, the relative difference of the energy differential decay rates obtained in these two gauges is smaller than $10^{-13}$. It is not a simple task to reach such high accuracy, especially in the low-$Z$ regime. The main reason for that is the standard {LAPACK} \cite{laug} routines which are commonly utilized to solve generalized eigenvalue problems and are designed to use double-precision arithmetic. They suffer, hence, a loss of numerical significance for large diagonally dominant matrices \cite{Amaro_2011}. In order to overcome this problem, we developed a novel $B$-spline code using the C++ template library Eigen \cite{eigenweb} which makes it possible to solve the generalized eigenvalue problem with arbitrary precision.

\begin{center}
\begin{figure}
\includegraphics[width=1.0\linewidth]{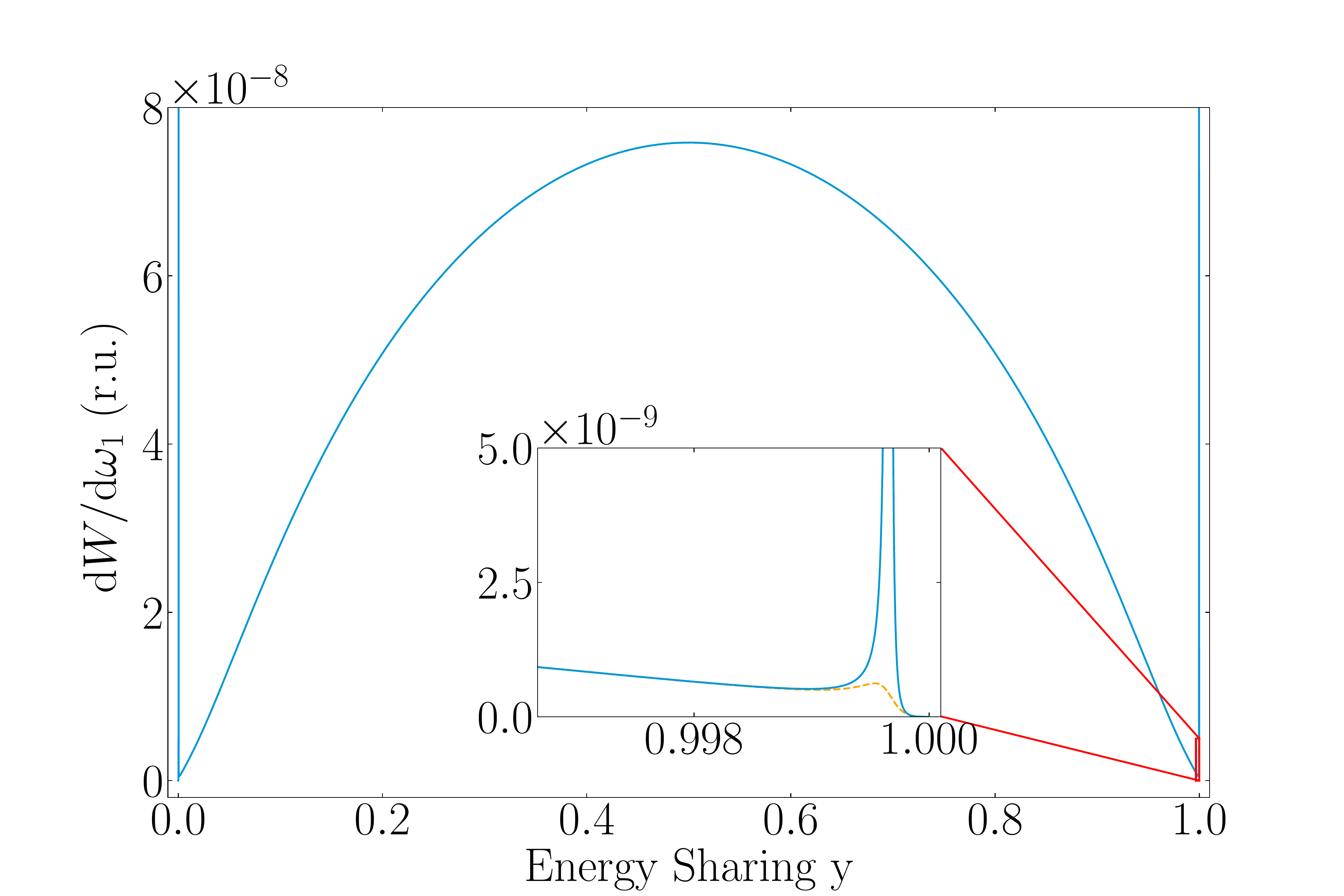}
\caption{Energy differential decay rate in relativistic units~(r.u.) for the $2s_{1/2} \to 1s_{1/2} + E1E1$ transition in hydrogen-like Uranium as a function of the energy sharing $y = \omega_1/(\omega_1 + \omega_2)$. The resonances close to the edges are either untreated (blue solid line) or rendered finite with the method by Labzowsky and co-workers~\cite{PhysRevA.80.062514} (orange dashed line).} \label{FinLW}
\end{figure}
\end{center}

When calculating the total decay rates, we face another difficulty that lowers the gauge invariance and hence accuracy of our calculations: The well-known problem of intermediate-state resonances. This problem does not arise for $2s_{1/2} \to 1s_{1/2}$ and $2p_{1/2} \to 1s_{1/2}$ transitions for point-like nuclei, because of the degeneracy of the $2s_{1/2}$ and $2p_{1/2}$ states. However, the degeneracy is lifted when dealing with finite nuclear size, which leads to the fact that the $2s_{1/2} \to 1s_{1/2}$ decay may proceed via the real intermediate $2p_{1/2}$ state. The existence of this intermediate state might be observed as a resonance in the energy differential decay rate, as seen from Fig. \ref{FinLW}. The appearance of such a resonance makes the integration over the photon energy and, hence, the calculation of the total rates cumbersome as was discussed in the literature~\cite{PhysRevA.77.042507,PhysRevA.80.062514}. In this work, we follow the approach proposed by Labzowsky and co-workers \cite{PhysRevA.80.062514} and introduce the finite linewidth of the resonance state to the electron propagator in a small interval around the divergent resonance. For example in Fig. \ref{FinLW}, this procedure is illustrated by the orange dashed line for the $2s_{1/2} \to 1s_{1/2} + E1E1$ transition in hydrogen-like Uranium. Of course, this method leads to the loss of gauge invariance in the small interval. However, since the resonances are located near the edges of the spectral distribution, for example at $y = \omega_1/(\omega_1+\omega_2)=0.00034$ and $y = 0.99966$ for U$^{91+}$, the violation of the gauge invariance does not affect the total rate significantly. The results of our calculations show that introducing an artificial linewidth to fix the resonance problem does not result in a violation of the gauge invariance above the level of $10^{-7}$ for the total rate. Therefore, all digits presented in the next chapter are still significant.

Yet another source of uncertainty of the two-photon calculations is imprecise knowledge about the nuclear size $R$. With the compilation of Angeli and Marinova \cite{ANGELI201369}, this leads to relative uncertainties in the calculation of the decay rates that do not exceed $10^{-8}$. We confirm this level of accuracy also when comparing calculations performed in the framework of the two nuclear models, namely the Fermi and solid sphere model.

Finally, to verify the correct implementation of the $B$-spline method and the electron-nucleus interaction potentials, we present in Table \ref{Eshift} the binding energies for the point-like nucleus and the energy shift due to finite nuclear size and vacuum polarization effects. These energies are obtained as the eigenvalues of the generalized eigenvalue problem (\ref{EigenValue}) and, as seen from the table, show a good agreement with the results from Yerokhin and Shabaev~\cite{doi:10.1063/1.4927487}.

\begin{table}
\begin{threeparttable}
\caption{Binding energies for a point-like nucleus and energy shifts due to finite nuclear size and vacuum polarization effects in units of eV. Results are presented for the $1s_{1/2}$, $2s_{1/2}$ and $2p_{1/2}$ states of hydrogen-like ions with nuclear charge $Z$. Brackets denote powers of 10.} \label{Eshift}
\begin{tabular}{cccc}
\hline
\hline
\vspace{0.1cm}
$Z$ & $E_\text{pnt}$ & $\Delta E_\text{fnt}$ & $\Delta E_\text{Uehl}$\\
\hline
\multicolumn{4}{c}{$1s_{1/2}$}\\
\makecell{1\\\hphantom{1}} & \makecell{-13.6059\hphantom{$^b$}\\-$13.6059^b$}& \makecell{4.99(-9)\hphantom{$^a$}\\4.99(-9)$^a$}& \makecell{-8.90(-7)\hphantom{$^a$}\\-8.90(-7)$^a$}\\
\makecell{40\\\hphantom{1}} & \makecell{-22253.68\hphantom{$^b$}\\-$22253.68^b$} &\makecell{0.516\hphantom{$^a$}\\$0.516^a$} & \makecell{-2.084\hphantom{$^a$}\\-$2.083^a$}\\
\makecell{92\\\hphantom{1}} & \makecell{-132279.93\hphantom{$^b$}\\-$132279.93^b$} & \makecell{$198.7$\hphantom{$^a$}\\$199.0^a$} & \makecell{-93.8\hphantom{$^a$}\\-$93.6^a$}\\
\multicolumn{4}{c}{$2s_{1/2}$}\\
\makecell{1\\\hphantom{1}} &\makecell{-3.4015\hphantom{$^b$}\\-$3.4015^b$}&\makecell{6.24(-10)\hphantom{$^a$}\\6.25(-10)$^a$}& \makecell{-1.11(-7)\hphantom{$^a$}\\-1.11(-7)$^a$} \\
\makecell{40\\\hphantom{1}} & \makecell{-5594.04\hphantom{$^b$}\\-$5594.04^b$} & \makecell{0.0696\hphantom{$^a$}\\$0.0696^a$}& \makecell{-0.277\hphantom{$^a$}\\-$0.277^a$}\\
\makecell{92\\\hphantom{1}} & \makecell{-34215.48\hphantom{$^b$}\\-$34215.48^b$}& \makecell{37.81\hphantom{$^a$}\\$37.71^a$} & \makecell{-16.50\hphantom{$^a$}\\-$16.46^a$}\\
\multicolumn{4}{c}{$2p_{1/2}$}\\
\makecell{1\\\hphantom{1}} & \makecell{-3.4015\hphantom{$^b$}\\-$3.4015^b$}& \makecell{0.\hphantom{$^a$}\\0.$^a$} & \makecell{-1.3(-12)\hphantom{$^a$}\\-1.3(-12)$^a$}\\
\makecell{40\\\hphantom{1}} & \makecell{-5594.04\hphantom{$^b$}\\-$5594.04^b$} & \makecell{0.0012\hphantom{$^a$}\\$0.0012^a$}& \makecell{-0.0068\hphantom{$^a$}\\-$0.0068^a$}\\
\makecell{92\\\hphantom{1}} & \makecell{-34215.48\hphantom{$^b$}\\-$34215.48^b$} & \makecell{4.41\hphantom{$^a$}\\$4.42^a$} & \makecell{-2.91\hphantom{$^a$}\\-$2.91^a$}\\
\hline
\hline
\end{tabular}
\begin{tablenotes}
\small 
\item $^a$Yerokhin and Shabaev~\cite{doi:10.1063/1.4927487}
\item $^b$Analytical formula from relativistic Dirac theory
\end{tablenotes}
\end{threeparttable}
\end{table}

\section{Results and Discussion} \label{ResADis}
\subsection{$2s_{1/2} \to 1s_{1/2}$ two-photon transition}
By using Eqs. (\ref{Matel2}) - (\ref{totalRate}) and the finite basis set method, discussed in Sec. \ref{finiteB}, we calculate the total rates for the two-photon decay of hydrogen-like ions. We start our discussion with the well-known $2s_{1/2} \to 1s_{1/2} + 2\gamma$ transition. In leading order, this decay proceeds with the emission of two electric dipole photons. The total rates $W_{E1,E1}$ for this $2E1$ transition are displayed in Table~\ref{E1E1} for nuclear charges ranging from $Z = 1$ to $Z = 92$ and for different electron-nucleus interaction potentials. In particular, calculations have been performed for the point-like and finite-size nucleus and by accounting the vacuum polarization as described by the Uehling potential (\ref{UehlPot}). Moreover, by following the well-known non-relativistic $Z$-behaviour \cite{KLARSFELD1969382}, we present our results divided by $Z^6$. For the naive point-like nuclear model, a good agreement with the previous calculations by Goldman \cite{PhysRevA.40.1185} and Filippin and co-workers~\cite{PhysRevA.93.012517} is obtained. Here, we agree up to the level of approximately $10^{-7}$ and $10^{-9}$, respectively. We note moreover, a misprint in the first digit of the value for $Z = 20$ of Goldman \cite{PhysRevA.40.1185}.

In order to investigate the effects of the finite nuclear size for the $2s_{1/2} \to 1s_{1/2} + 2E1$ decay rates, we performed calculations also for the hard-sphere and Fermi distribution. As mentioned already above, both models agree on the level of $10^{-8}$ and are presented in the third column of Table \ref{E1E1}. As seen from the table, the finite nuclear size reduces the decay rates by $2.79 \cdot 10^{-8}$\% for neutral Hydrogen and up to 0.0092\% for hydrogen-like Uranium ions which is in good agreement with the findings from Parpia and Johnson \cite{PhysRevA.26.1142} and Labzwosky and co-workers \cite{Labzowsky_2005} within their estimated error. In order to understand this reduction, we have to revisit the decay rate (\ref{rateMult}). As one can see from this expression, the finite nuclear size may affect both, the second-order matrix element $\widetilde{M}_{fi}$ and the energy prefactor $\omega_1\omega_2$. Looking again at Table \ref{Eshift}, we can see that the finite nuclear size leads to the reduction of the transition energy $\omega_1+\omega_2$ which can reach $0.16\%$ for U$^{91+}$. In contrast, our calculations indicate that the two-photon matrix element increases if, in place of the point-like, the finite-sized nucleus is used. As a result, the reduction of the transition energy and the enhancement of $\widetilde{M}_{fi}$ leads to the fact that the total $2s_{1/2}\to1s_{1/2} + 2\gamma$ decay rate is affected by the finite nuclear size effect less than one would expect from analysing the transition energy, only.

Yet another correction that may affect the $2s_{1/2} \to 1s_{1/2} + 2\gamma$ decay rate is the vacuum polarization which is approximated by the Uehling potential. As seen from the fourth column of Table \ref{E1E1}, this correction increases the decay rate by $4.96\cdot10^{-6}\%$ for Hydrogen and 0.03\% for U$^{91+}$. By comparing the third and fourth column of Table \ref{E1E1}, we see that the vacuum polarization correction to the total decay rate is always larger than the one obtained from the finite nuclear size. This is especially surprising since the transition energies are known to be stronger influenced by the finite nuclear size in the high-$Z$ regime as also seen from Table \ref{Eshift}. To explain this unexpected behaviour, we have to look again at Eq. (\ref{rateMult}). As already explained above, for the $2s_{1/2}\to 1s_{1/2}+2E1$ transition, the finite nuclear size reduces the transition energies and enhances the matrix elements. In contrast, the vacuum polarization acts just in the opposite direction and, similar to the finite nuclear size correction, the QED contributions to the matrix element $\widetilde{M}_{fi}$ and energy prefactor partially cancel each other. However, this cancellation is less pronounced for the vacuum polarization case, thus, resulting in the stronger sensitivity of the results to the Uehling correction.

\begin{table}
\begin{threeparttable}
\caption{Total two-photon $2s_{1/2}\to 1s_{1/2}$ decay rates for hydrogen-like ions with nuclear charge $Z$ in units $Z^6$s$^{-1}$. Calculations have been performed for the $E1E1$ multipole channel and for a point-like nucleus ($V_\text{pnt}$), a finite-sized nucleus ($V_\text{fnt}$) and a finite nucleus with additional Uehling potential ($V_\text{fnt} + V_\text{Uehl}$).} \label{E1E1}
\begin{tabular}{ccccc}
\hline
\hline
$Z$ & $V_\text{pnt}$ & $V_\text{fnt}$ & $V_\text{fnt} + V_\text{Uehl}$\\
\hline
\rule{0mm}{0.8cm}
\makecell{1\\\hphantom{1}\\\hphantom{1}} & \makecell{8.2290615\hphantom{$^a$}\\$8.2290615^a$\\$8.2290626^b$} & \makecell{8.2290615\\$8.22906^c$\hphantom{1}\\\hphantom{1}} & \makecell{8.2290619\\\hphantom{1}\\\hphantom{1}} &\\
\vspace{0.1cm}
\makecell{20\\\hphantom{1}} & \makecell{8.1174024\hphantom{$^a$}\\$9.1174035^{b}$} & \makecell{8.1173852\\$8.1095^c$\hphantom{11}} & \makecell{8.1175410\\\hphantom{1}}&\\
\vspace{0.1cm}
\makecell{40\\\hphantom{1}\\\hphantom{1}} & \makecell{7.8092601\hphantom{$^a$}\\$7.8092601^a$\\$7.8092612^b$} & \makecell{7.8091196\\$7.8013^c$\hphantom{11}\\\hphantom{1}} & \makecell{7.8097303\\\hphantom{1}\\\hphantom{1}}&\\
\vspace{0.1cm}
\makecell{60\\\hphantom{1}} & \makecell{7.3446473\hphantom{$^a$}\\$7.3446482^b$} & \makecell{7.344077\hphantom{1}\\$7.3365^c$\hphantom{11}} & \makecell{7.3453658\\\hphantom{1}}&\\
\vspace{0.1cm}
\makecell{80\\\hphantom{1}} & \makecell{6.7428868\hphantom{$^a$}\\$6.7428876^b$} & \makecell{6.74157\hphantom{11}\\$6.7348^c$\hphantom{11}} & \makecell{6.74347\\\hphantom{1}}\hphantom{11}&\\
\vspace{0.1cm}
\makecell{92\\\hphantom{1}\\\hphantom{1}} & \makecell{6.3096615\hphantom{$^a$}\\$6.3096618^a$\\$6.3096623^b$} & \makecell{6.30908\hphantom{11}\\$6.3026^c$\hphantom{11}\\\hphantom{1}} & \makecell{6.31098\hphantom{11}\\\hphantom{1}\\\hphantom{1}}&\\
\hline
\hline
\end{tabular}
\begin{tablenotes}
\small 
\item $^a$Filippin and co-workers~\cite{PhysRevA.93.012517}
\item $^b$Goldman~\cite{PhysRevA.40.1185}, there seems to be a misprint in the value for $Z = 20$
\item $^c$Labzowsky and co-workers~\cite{Labzowsky_2005}
\end{tablenotes}
\end{threeparttable}
\end{table}

So far, we discussed the leading $2E1$ channel of the $2s_{1/2} \to 1s_{1/2} + 2\gamma$ transition. Owing to the selection rules, the other -- much weaker -- channels can also contribute to this decay. In order to investigate the role of the higher-order multipole contributions, we calculated the total decay rate $W = \sum_{p_1L_1p_2L_2} W(p_1L_1,p_2L_2)$ where the summation over $L_1$ and $L_2$ runs up to $L_\text{max}$ = 4. The results of these calculations are shown in Table \ref{2sT1s} again for a point nucleus and including the finite nuclear size and vacuum polarization corrections. As seen from the table, the contributions of both corrections are qualitatively similar to what was observed for the leading $2E1$ channel. In particular, the total two-photon decay rates are reduced by the finite nuclear size and enhanced if the vacuum polarization is taken into account.

\begin{table}
\begin{threeparttable}
\caption{Total two-photon $2s_{1/2}\to 1s_{1/2}$ decay rates for hydrogen-like ions with nuclear charge $Z$ in units $Z^6$s$^{-1}$. Results have been obtained for a point-like nucleus ($V_\text{point}$), a finitely sized nucleus ($V_\text{finite}$) and a finite nucleus with additional Uehling potential ($V_\text{fnt} + V_\text{Uehl}$). Moreover, we have performed the summation over all allowed multipole channels $p_1L_1,p_2L_2$ for $L_1,L_2 = 1...4$.} \label{2sT1s}
\begin{tabular}{ccccc}
\hline
\hline
Z & $V_\text{point}$ & $V_\text{finite}$ & $V_\text{fnt} + V_\text{Uehl}$ &\\
\hline
\vspace{0.1cm}
\rule{0mm}{0.8cm}
\makecell{1\\\hphantom{1}\\\hphantom{1}} & \makecell{8.2290615\hphantom{$^a$}\\$8.2290615^a$\\$8.2290626^b$} & \makecell{8.2290615\\\hphantom{1}\\\hphantom{1}} & \makecell{8.2290619\\\hphantom{1}\\\hphantom{1}} & \\
\vspace{0.1cm}
\makecell{20\\\hphantom{1}\\\hphantom{1}} & \makecell{8.1174454\hphantom{$^a$}\\$8.1174464^a$\\$8.1174466^b$} & \makecell{8.1174282\\\hphantom{1}\\\hphantom{1}} & \makecell{8.1175840\\\hphantom{1}\\\hphantom{1}} &\\
\vspace{0.1cm}
\makecell{40\\\hphantom{1}\\\hphantom{1}} & \makecell{7.8099289\hphantom{$^a$}\\$7.8099289^a$\\$7.8099299^b$} & \makecell{7.8097883\\\hphantom{1}\\\hphantom{1}} & \makecell{7.8103993\\\hphantom{1}\\\hphantom{1}} &\\
\vspace{0.1cm}
\makecell{60\\\hphantom{1}\\\hphantom{1}} & \makecell{7.3479098\hphantom{$^a$}\\$7.3479098^a$\\$7.3479109^b$} & \makecell{7.347339\hphantom{1}\\\hphantom{1}\\\hphantom{1}} & \makecell{7.3486296\\\hphantom{1}\\\hphantom{1}} &\\
\vspace{0.1cm}
\makecell{80\\\hphantom{1}\\\hphantom{1}} & \makecell{6.7528665\hphantom{$^a$}\\$6.7528660^a$\\$6.7528675^b$} & \makecell{6.75154\hphantom{11}\\\hphantom{1}\\\hphantom{1}} & \makecell{6.75345\hphantom{11}\\\hphantom{1}\\\hphantom{1}} &\\
\vspace{0.1cm}
\makecell{92\\\hphantom{1}\\\hphantom{1}} & \makecell{6.3269332\hphantom{$^a$}\\$6.326931^b$\hphantom{1}\\$6.3269340^a$} & \makecell{6.32633\hphantom{11}\\\hphantom{1}\\\hphantom{1}} & \makecell{6.32821\hphantom{11}\\\hphantom{1}\\\hphantom{1}} &\\
\hline
\hline
\end{tabular}
\begin{tablenotes}
\small 
\item $^a$Filippin and co-workers~\cite{PhysRevA.93.012517}
\item $^b$Goldman~\cite{PhysRevA.40.1185}
\end{tablenotes}
\end{threeparttable}
\end{table}

\subsection{$2p_{1/2} \to 1s_{1/2}$ two-photon transition}
Besides the well-known $2s_{1/2}  \to 1s_{1/2} + 2\gamma$ decay, we also apply our theory to discuss the $2p_{1/2} \to 1s_{1/2} + 2\gamma$ transition. In the leading order, this decay may proceed via $E1M1$ or $E1E2$ channels. The results for the $E1M1$ channel are presented in Table \ref{E1M1}, again for a point-like and finite-size nucleus as well as accounting for the Uehling correction. The decay rates are obtained in the range from $Z=1$ to $Z=92$ and are rescaled as $Z^8$ as suggested by the non-relativistic limit \cite{Labzowsky_2005}. Similar to before, we start the discussion of our results from the finite nuclear size correction. As seen from the table, this correction reduces the decay rate as it was also observed for $2s_{1/2}  \to 1s_{1/2} + 2\gamma$ transition. In contrast to that transition, however, the reduction of the $2p_{1/2}\to 1s_{1/2}+E1M1$ decay rate is significantly stronger ranging from $1.34\cdot10^{-7}$\% for Hydrogen to 0.352\% for hydrogen-like Uranium. Once more, this can be explained by the analysis of the finite nuclear size corrections to the matrix elements and the transition energies in Eq. (\ref{rateMult}). For the $2p_{1/2}\to 1s_{1/2} + 2\gamma$ case, both get reduced if, in place of the point-like, the finite-sized nucleus is considered. These reductions, therefore, reinforce each other which results in the larger correction.

As seen from the fourth column of Table \ref{E1M1}, the Uehling contribution increases the decay rates between $2.39\cdot10^{-5}\%$ and 0.187\%. The overall larger effect, compared to the $2s_{1/2} \to 1s_{1/2} + 2\gamma$ transition, can be again explained by the reinforcement of the QED contributions on the transition energies and matrix elements.

\begin{table}
\begin{threeparttable}
\caption{Total two-photon $2p_{1/2}\to 1s_{1/2}$ decay rates for hydrogen-like ions with nuclear charge $Z$ in units $Z^{8}10^{-6}$s$^{-1}$. Calculations have been performed for the $E1M1$ multipole channel and for a point-like nucleus ($V_\text{pnt}$), a finite-sized nucleus ($V_\text{fnt}$) and a finite nucleus with additional Uehling potential ($V_\text{fnt} + V_\text{Uehl}$).} \label{E1M1}
\begin{tabular}{ccccc}
\hline
\hline
Z & $V_\text{pnt}$ & $V_\text{fnt}$ & $V_\text{fnt} + V_\text{Uehl}$\\
\hline
\rule{0mm}{0.6cm}
\makecell{1\\\hphantom{1}} & \makecell{9.6766569\hphantom{$^a$}\\$9.6777569^a$} & \makecell{9.6766569\\$9.667^b$\hphantom{111}} & \makecell{9.6766592\\\hphantom{1}}\\

\makecell{20\\\hphantom{1}} & \makecell{9.5561970\hphantom{$^a$}\\\hphantom{1}} & \makecell{9.5561068\\$9.543^b$\hphantom{111}} & \makecell{9.5569117\\\hphantom{1}}\\

\makecell{40\\\hphantom{1}} & \makecell{9.1973052\hphantom{$^a$}\\$9.1973052^a$} & \makecell{9.1966168\\$9.186^b$\hphantom{111}} & \makecell{9.1994459\\\hphantom{1}}\\

60 & 8.6260732\hphantom{$^a$} & 8.6231951 & \makecell{8.6288466\\\hphantom{1}}\\

\makecell{80\\\hphantom{1}} & \makecell{7.9316051\hphantom{$^a$}\\\hphantom{1}} & \makecell{7.9211520\\$7.910^b$\hphantom{111}} & \makecell{7.9307704\\\hphantom{1}}\\

\makecell{92\\\hphantom{1}} & \makecell{7.5541404\hphantom{$^a$}\\$7.5541404^a$} & \makecell{7.5275515\\$7.519^b$\hphantom{111}} & \makecell{7.5416695\\\hphantom{1}}\\
\hline
\hline
\end{tabular}
\begin{tablenotes}
\small 
\item $^a$Filippin and co-workers~\cite{PhysRevA.93.012517}
\item $^b$Labzowsky and co-workers~\cite{Labzowsky_2005}
\end{tablenotes}
\end{threeparttable}
\end{table}

The other leading channel that contributes to the $2p_{1/2}\to 1s_{1/2}+2\gamma$ decay is the $E1E2$ transition for which the total rates are shown in Table \ref{E1E2}. This multipole transition exhibits the same $Z$-scaling law and also generally the same behaviour with respect to the corrections due to the finite-size nucleus and vacuum polarization as it was observed for the $E1M1$ channel. For the $E1E2$ channel, however, the decay is even stronger affected by the finite nuclear size and QED effects which may reach, for the decay of hydrogen-like Uranium, a reduction by $0.692\%$ and enhancement by $0.322\%$, respectively.

\begin{table}
\begin{threeparttable}
\caption{Total two-photon $2p_{1/2}\to 1s_{1/2}$ decay rates for hydrogen-like ions with nuclear charge $Z$ in units $Z^{8}10^{-6}$s$^{-1}$. Calculations have been performed for the $E1E2$ multipole channel and for a point-like nucleus ($V_\text{pnt}$), a finite-sized nucleus ($V_\text{fnt}$) and a finite nucleus with additional Uehling potential ($V_\text{fnt} + V_\text{Uehl}$).}  \label{E1E2}
\begin{tabular}{ccccc}
\hline
\hline
Z & $V_\text{pnt}$ & $V_\text{fnt}$ & $V_\text{fnt} + V_\text{Uehl}$\\
\hline
\rule{0mm}{0.6cm}
\makecell{1\\\hphantom{1}} & \makecell{6.6117981\hphantom{$^a$}\\$6.6117981^a$} & \makecell{6.6117981\\$6.605^b$\hphantom{111}} & \makecell{6.6118003\\\hphantom{1}}\\
\makecell{20\\\hphantom{1}} & \makecell{6.5202286\hphantom{$^a$}\\\hphantom{1}} & \makecell{6.5201386\\$6.516^b$\hphantom{111}} & \makecell{6.5209400\\\hphantom{1}}\\
\makecell{40\\\hphantom{1}} & \makecell{6.2446748\hphantom{$^a$}\\$6.2446748^a$} & \makecell{6.2439316\\$6.238^b$\hphantom{111}} & \makecell{6.2469332\\\hphantom{1}}\\
60 & 5.7832261\hphantom{$^a$} & 5.7797060 & 5.7862409\\
\makecell{80\\\hphantom{1}} & \makecell{5.1262923\hphantom{$^a$}\\\hphantom{1}} & \makecell{5.1123097\\$5.107^b$\hphantom{111}} & \makecell{5.1237402\\\hphantom{1}}\\
\makecell{92\\\hphantom{1}} & \makecell{4.6272865\hphantom{$^a$}\\$4.6272865^a$} & \makecell{4.5952682\\$4.591^b$\hphantom{111}} & \makecell{4.6101315\\\hphantom{1}}\\
\hline
\hline
\end{tabular}
\begin{tablenotes}
\small 
\item $^a$Filippin and co-workers~\cite{PhysRevA.93.012517}
\item $^b$Labzowsky and co-workers~\cite{Labzowsky_2005}
\end{tablenotes}
\end{threeparttable}
\end{table}

Similar to before, we also investigate the total $2p_{1/2} \to 1s_{1/2} + 2\gamma$ decay rate by including higher-order multipole terms up to $L_\text{max}=4$. The total rates for the point-like and finite nucleus as well as for the additional Uehling correction are shown in Table \ref{2pT1s}. As seen from the table, these corrections generally show the same behaviour as for the $E1M1$ and $E1E2$ channel. For U$^{91+}$ ions, they lead to a reduction of the total rate by $0.484\%$ due to the finite nuclear size and enhancement by $0.239\%$ due to the vacuum polarization.

\begin{table}
\begin{threeparttable}
\caption{Total two-photon $2p_{1/2}\to 1s_{1/2}$ decay rates for hydrogen-like ions with nuclear charge $Z$ in units $Z^810^{-6}$s$^{-1}$. Results have been obtained for a point-like nucleus ($V_\text{pnt}$), a finitely sized nucleus ($V_\text{fnt}$) and a finite nucleus with additional Uehling potential ($V_\text{fnt} + V_\text{Uehl}$). Moreover, we have performed the summation over all allowed multipole channels $p_1L_1,p_2L_2$ for $L_1,L_2 = 1...4$.} \label{2pT1s}
\begin{tabular}{ccccc}
\hline
\hline
Z & $V_\text{pnt}$ & $V_\text{fnt}$ & $V_\text{fnt} + V_\text{Uehl}$\\
\hline
\rule{0mm}{0.6cm}
\makecell{1\\\hphantom{1}} & \makecell{16.288455\hphantom{$^a$}\\16.288455$^a$} & \makecell{16.288455\\\hphantom{1}} & \makecell{16.288460\\\hphantom{1}}\\

20 & 16.076447\hphantom{$^a$} & 16.076267 & 16.077873\\

\makecell{40\\\hphantom{1}} & \makecell{15.442308\hphantom{$^a$}\\$15.442308^a$} & \makecell{15.440876\\\hphantom{1}} & \makecell{15.446707\\\hphantom{1}}\\

60 & 14.410849\hphantom{$^a$} & 14.404450 & 14.416639\\

80 & 13.062372\hphantom{$^a$} & 13.037919 & 13.058982\\

\makecell{92\\\hphantom{1}} & \makecell{12.188751\hphantom{$^a$}\\$12.188750^a$} & \makecell{12.130067\\\hphantom{1}} & \makecell{12.159084\\\hphantom{1}}\\
\hline
\hline
\end{tabular}
\begin{tablenotes}
\small 
\item $^a$Filippin and co-workers~\cite{PhysRevA.93.012517}
\end{tablenotes}
\end{threeparttable}
\end{table}

\section{Summary and Outlook} \label{SumAOut}
In conclusion, we presented a theoretical study of the two-photon decay of hydrogen-like ions. Special attention was paid to the total rates obtained upon integration over the energies and directions of the emitted photons. In order to calculate these total decay rates, we have used the framework of quantum electrodynamics. The use of QED theory has allowed us to naturally account for the finite nuclear size and vacuum polarization corrections. In order to study the size of these corrections, high precision calculations have been performed for the $2s_{1/2} \to 1s_{1/2} + 2\gamma$ and $2p_{1/2} \to 1s_{1/2} + 2\gamma$ two-photon decay of hydrogen-like ions in the range from $Z=1$ to $Z=92$. Our results have shown that these two well-established transitions are affected differently by the QED and finite nuclear size effects. Particularly, both effects are more pronounced for the $2p_{1/2}\to 1s_{1/2} + 2\gamma$ case. For example, the transition rate in U$^{91+}$ is reduced by $0.484\%$ by the finite nuclear size and enhanced by $0.239\%$ by the Uehling correction. For the $2s_{1/2} \to 1s_{1/2} + 2\gamma$ decay of hydrogen-like Uranium, in contrast, these two corrections are just $0.0092\%$ and $0.03\%$, respectively.

The finite nuclear size and QED corrections discussed in this manuscript are too small in order to be observed in present-day two-photon experiments. They might be significantly increased, however, when replacing the usual electronic ions by muonic systems. The high-precision $B$-spline approach developed in the present work can be easily extended to the transition in such muonic ions. The study of two-photon decay of muonic ions is currently under development and its results will be published in a future work.

\begin{acknowledgments}
This work has been supported by the GSI Helmholtz Centre for Heavy Ion Research under the project BSSURZ1922 and by the DFG under the project SU658/4-1. 
\end{acknowledgments}

\end{document}